\documentclass[
    ,final            
  ]
  {aipproc}

\layoutstyle{6x9}

\begin{document}

\title{The Pierre Auger Project and Enhancements}

\author{A. Etchegoyen}{
  address={ITeDA, Instituto de Tecnologías en Detección y Astropartículas (CNEA, CONICET,
  UNSAM), Argentina},altaddress={UTN - FRBA, Argentina}
}

\author{U. Fr\"ohlich}{
 address={Universit\"at Siegen, Siegen, Germany} }

\author{A. Lucero}{
 address={ITeDA, Instituto de Tecnologías en Detección y Astropartículas (CNEA, CONICET,
  UNSAM), Argentina}
}

\author{I. Sidelnik}{
 address={ITeDA, Instituto de Tecnologías en Detección y Astropartículas (CNEA, CONICET,
  UNSAM), Argentina}
}

\author{B. Wundheiler}{
 address={ITeDA, Instituto de Tecnologías en Detección y Astropartículas (CNEA, CONICET,
  UNSAM), Argentina}
}
\author{for the Pierre Auger Collaboration}{
 address={Observatorio Pierre Auger, Av. San Mart\'in Norte 304, 5613 Malarg\"ue, Argentina} }

\begin{abstract}
The current status of the scientific results of the Auger
Observatory will be discussed which include spectrum, anisotropy
in arrival directions, chemical composition analyses, and limits
on neutrino and photon fluxes. A review of the Observatory
detection systems will be presented. Auger has started the
construction of its second phase which encompasses antennae for
radio detection of cosmic rays, high-elevation telescopes, and
surface plus muon detectors. Details will be presented on the
latter, AMIGA (\textit{Auger Muons and Infill for the Ground
Array}), an Auger project consisting of 85 detector pairs each one
composed of a surface water-Cherenkov detector and a buried muon
counter. The detector pairs are arranged in an array with spacings
of 433 and 750 m in order to perform a detailed study of the
10$^{17}$ eV to 10$^{19}$ eV spectrum region. Preliminary results
on the performance of the 750 m array of surface detectors and the
first muon counter prototype will be presented.

\end{abstract}

\maketitle

\section{Introduction}

Cosmic ray observatories aim to experimentally cast light on the
energy, origin, and chemical composition of the primary particles
arriving to the top of the Earth atmosphere. They would generally
consist on surface detectors deployed over a given area with a
chosen geometry in order to reconstruct the shower lateral
distribution at earth's surface (e.g. the observatory at Volcano
Ranch, New Mexico where in 1962 the first airshower with ascribed
energy in excess of 100 EeV \cite{Linsley:1980} was recorded). A
different technique based on optical telescopes was developed and
implemented by Fly's Eye/HiRes \cite{Baltrusaitis:1985}. They
reconstruct the longitudinal shower profile by detecting the
fluorescence light produced by excitation of atmospheric nitrogen
by shower electrons. Surface detectors have a 100\% duty cycle and
make use of simulations in order to evaluate the primary energy,
while the fluorescence technique has a 10 - 15\% duty cycle
(clear, dark nights) and does not resort to simulations for the
energy estimation, although it relies on the atmospheric
fluorescence yield, on a continuous evaluation of the atmospheric
light attenuation length and on the absolute calibration of the
telescopes. The aperture is well defined for the surface array at
higher energies while for the telescopes it increases with energy
and needs to be calculated with Monte Carlo simulations. They are
clearly two systems that complement each other and for this reason
the Auger Observatory uses a hybrid technique with both a surface
detector (SD) array and fluorescence detector (FD) system.

As above mentioned, it is of paramount importance to measure the
energy of the arriving cosmic rays. With this measurement it is
possible to obtain the cosmic ray spectrum, i.e. the flux of
primary particles as a function of their incoming energy.

\section{The Cosmic Ray Spectrum}
\begin{figure}[!h]
\includegraphics[width=.40\textwidth]{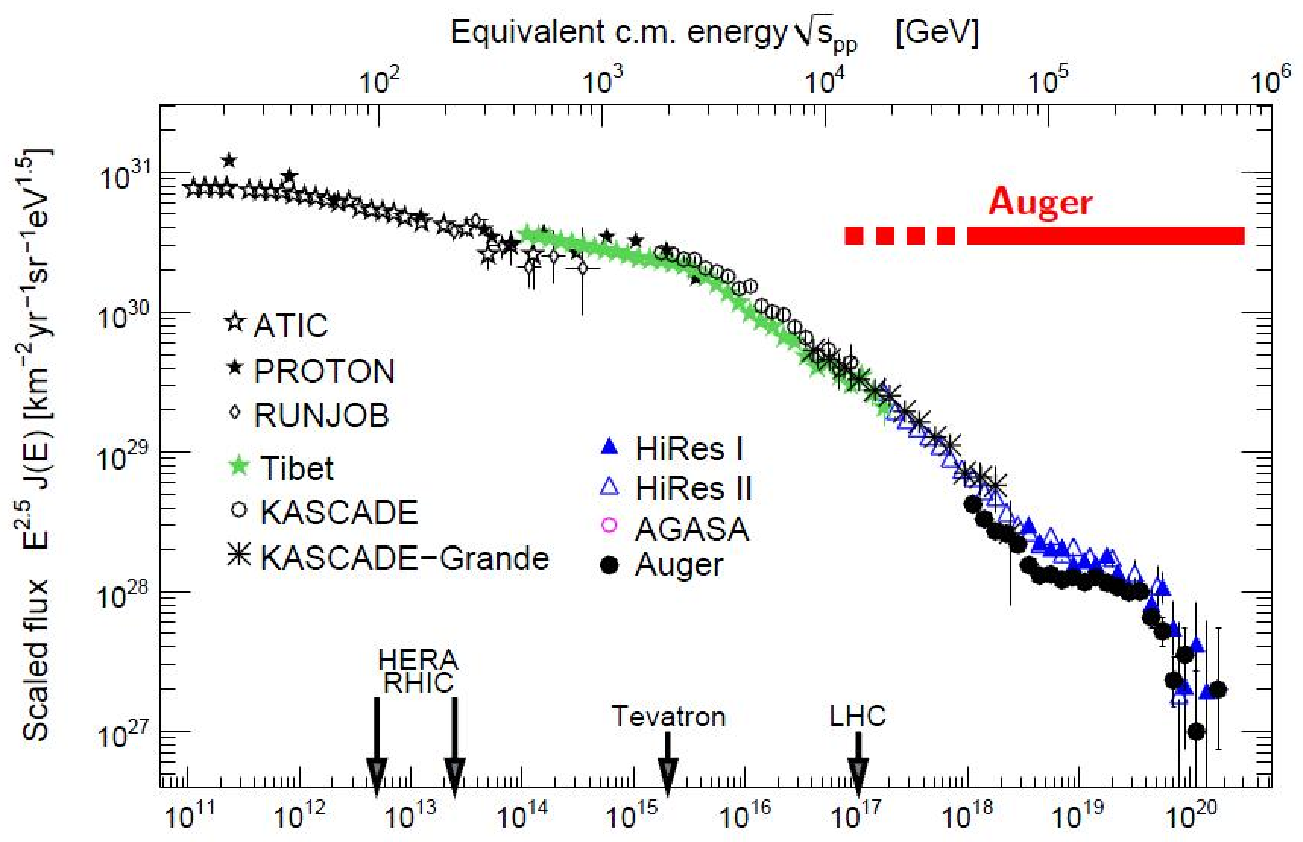}
\includegraphics[width=.40\textwidth]{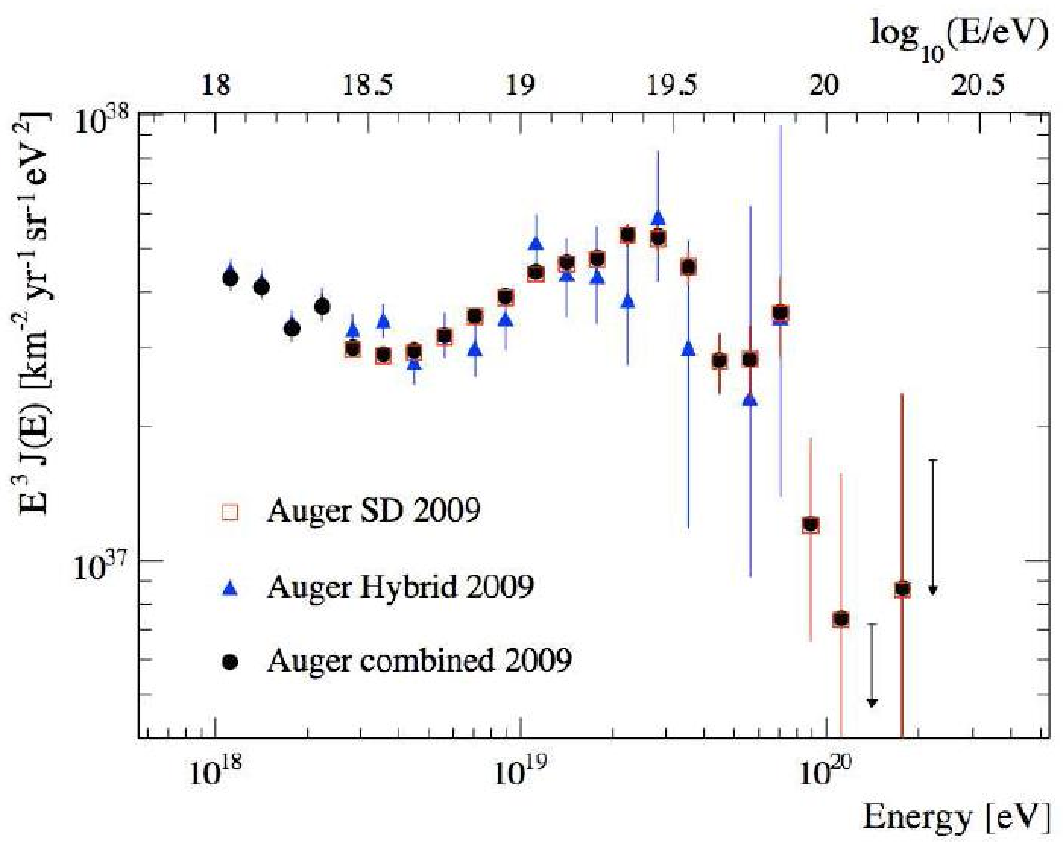}
\caption{The cosmic ray spectrum; \emph{(left)} compendium of data
from different observatories with an equivalent \emph{p-p}
collider-accelerator energy scale; \emph{(right)} Auger
Observatory spectrum from SD, hybrid, and combined data
\cite{Auger:Spectrum}. Note that the flux is multiplied by
E$^{3}$}\label{spectrum}
\end{figure}

The cosmic ray spectrum at higher energy only presents four
distinct traits, the ``knee'', the ``second knee'', the ``ankle'',
and the GZK cutoff and therefore a thorough understanding of
cosmic rays encompasses the study of these features including
their chemical composition. The knee occurs at $ 3-5 \times
10^{15} eV$ where the spectral index changes from -2.7 to -3.1
(see Fig. \ref{spectrum}.left), the second knee at $\sim$ 0.4 EeV
(1.0 EeV = $10^{18}$ eV) with a further steepening of the spectrum
(see Fig. \ref{spectrum}.left), and the ankle at $\sim$ 4 EeV (see
Fig. \ref{spectrum}.right). The GZK-cutoff (named after Greisen,
Zatsepin, and Kuz'min who suggested it) is a suppression of the
cosmic ray flux at very high energies \cite{Gresien:1966,
Zatsepin:1966} (see Fig. \ref{spectrum}.right) by interactions
with the microwave background radiation.

The second knee was observed by AKENO \cite{akeno}, Fly's Eye
stereo \cite{fleye1}, and its physical interpretation is still
uncertain. It might be just due to attaining the maximum cosmic
rays energies within the galaxy \cite{HAU05} since when this limit
is reached the flux will necessarily decay with a larger spectral
index. Or even more, it could be the transition region from
galactic to extragalactic primaries where the predominant
contribution arises from extragalactic proton primaries
\cite{BER05a,BER05b} due to the decreasing galactic flux.

The ankle was observed by HiRes \cite{HiRes:2009} at $\sim$ 3 EeV
and by AGASA \cite{AGASA:2003} but at a higher energy, $\sim$ 10
EeV . There are two main physical interpretations of the ankle
depending on where the transition from galactic to extragalactic
sources takes place. If the transition occurs at the second knee,
then the ankle (or the dip in the spectrum) is reported to arise
from $e^{-}e^{+}$ pair production from extragalactic protons
collisions with the microwave cosmic background radiation
\cite{BER05a,BER05b}. On the other hand, if the transition occurs
at the ankle, then the ankle would arise from the different
spectrum indexes of the galactic and extragalactic components
\cite{HILL04, WIB05, ALL05}.

The red horizontal line in Fig. \ref{spectrum}.left shows the
energy range of the Auger Observatory, the full line spans the
current range and the dashed line the range after the enhancements
which are currently being installed. The most relevant scientific
results of Pierre Auger Observatory will be presented after a
short outline of the Observatory detection systems that will help
to assess the data quality. Finally the current Observatory
upgrades and their prospective science will be described.

\section{The Pierre Auger Observatory}

The Pierre Auger Project \cite{Watson:2004} studies the highest
energies cosmic rays arriving on the surface of the earth from
outer space. It aims at building two Observatories situated in
both hemispheres and in November 2008 the austral observatory in
Malarg\"{u}e, Province of Mendoza, Argentina was formally
inaugurated. The Collaboration now focusses on the construction of
Auger North \cite{Bluemer:2010, Nitz:2010} in Colorado, USA, which
is designed to have a much larger acceptance in order to
significantly enhance the study cosmic rays with energies above
$\sim$ 60 EeV.

The Auger observatory has two distinctive features: its
exceptional size and its hybrid nature (i.e. both surface detector
\cite{Allekotte:2008} and fluorescence \cite{AugerFD:2010}
telescope systems). As such, Auger provides a large number of
events with better controlled systematic detection uncertainties.

The SD array consists of 1600 cylindrical water Cherenkov
detectors of 10 m$^{2} \times$ 1.2 m high arranged on a triangular
grid with 1.5 km spacing, covering an area of 3000 km$^{2}$ at
$\sim$ 1400 m above sea level. The principle of operation of these
detectors is that charged particles produce Cherenkov light when
traversing the 12 tons of pure water lodged in the tank. This
light is partially collected by three symmetrically placed 9"
photomultiplier tubes (PMT) at the water surface, placed 1.2 m
away from the tank center.

\begin{figure}[!h]
\includegraphics[height=.2\textheight]{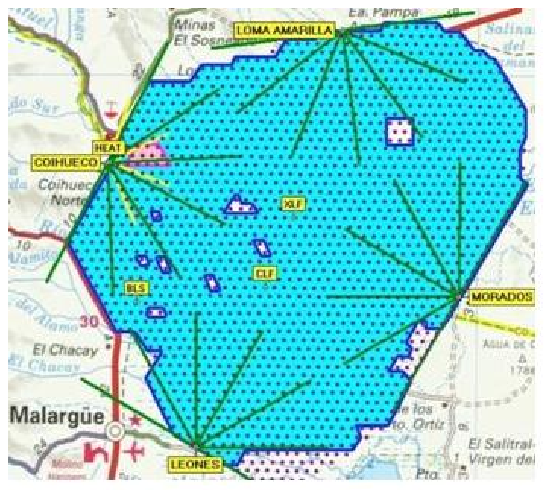}
\includegraphics [height=.2\textheight]{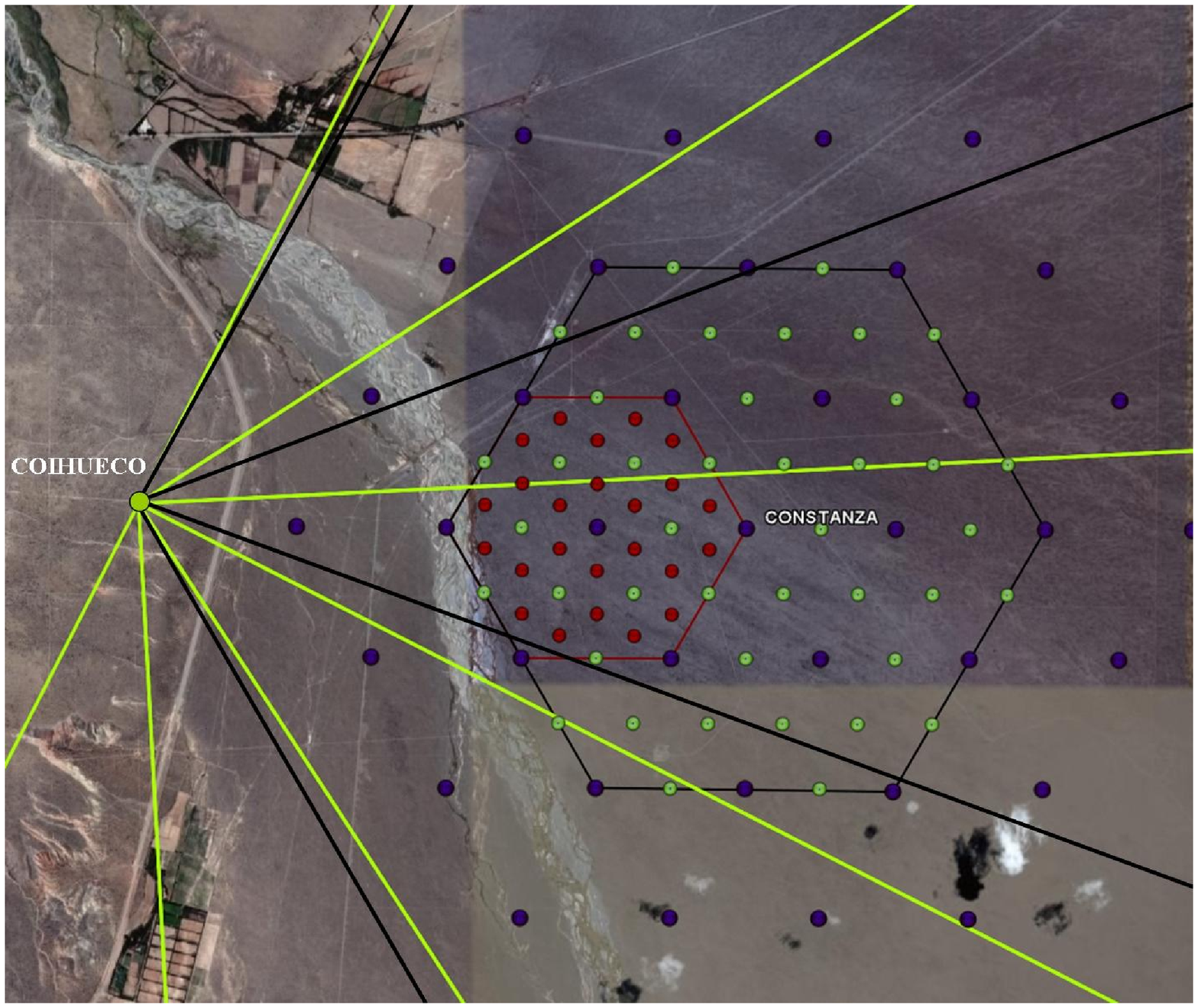}
\caption{Layout of the Auger Observatory, green lines limit the
FOV of the 6 telescopes with $1.5^{\circ }-30^{\circ }\times
0^{\circ }-30^{\circ}$ elevation and azimuth FOV. \emph{(left)}
Dots symbolize SD positions, those in the blue shaded area are
operational; \emph{(right)} layout of the Auger Observatory
upgrades near Cerro Coihueco. Black lines limit the $30^{\circ
}-58^{\circ }\times 0^{\circ }-30^{\circ}$ FOV for the 3 HEAT
telescopes. The two hexagons limit the AMIGA infilled areas of 5.9
and 23.5 km$^{2}$ with 433 and 750 m triangular grid detector
spacings, respectively. Each dot within these hexagons represents
a pair of a water Cherenkov detector and a muon counter. The
center dot, named Constanza, is placed $\sim$ 6.0 km away from
Cerro Coihueco. \label{Auger-map}}
\end{figure}

The FD system was originally designed to be constituted by 24
telescopes, deployed in 4 buildings hosting 6 telescopes each
overlooking the SD array (see Fig. \ref{Auger-map}.left). Each
telescope has an optical filter and corrector ring at the entrance
window, a mirror and a PMT camera on its focal plane where the
fluorescence light is collected. This system produces a light spot
with a 0.5$^{\circ}$ spread while each PMT has a FOV (Field of
View) of 1.5$^{\circ}$.

As mentioned already, the Observatory aims to study the energy,
origin and chemical composition of the primary particle. The
origin is studied by extrapolating back the reconstructed arrival
directions and the composition by measuring the atmospheric depth
where showers have their maximum development, X$_{max}$
\cite{AugerXmax:2010}. The energy calibration is dominated by FD
systematic uncertainties with a total estimated value of 22 \%
\cite{Giulio:2009}, the arrival direction uncertainty is better
than 1.0\% for primaries with energy above 10 EeV
\cite{AugerAD:2009}, and the X$_{max}$ uncertainty is 20
g/cm$^{2}$ \cite{AugerXmax:2010}.

\section{Scientific Results of the Pierre Auger
Observatory}

The most prominent scientific results obtained by the Pierre Auger
Observatory will now be outlined.

\paragraph{End of the Spectrum}
The cosmic ray flux was measured with both hybrid and SD array
data \cite{Auger:Spectrum} showing a good agreement. A combined
spectrum has been derived with high statistics covering the energy
range from 1.0 to above 100 EeV (see Fig. \ref{spectrum}.right)
with an estimated energy systematic uncertainty  of 22\%. The
ankle was found at $log_{10}(E_{ankle}/eV) = 18.61 \pm 0.01$. The
spectrum is suppressed at $log_{10}(E_{1/2}/eV) = 19.61 \pm 0.03$
with a significance in excess of 20 $\sigma$.

\paragraph{Anisotropy}
The last update on anisotropy data of correlations between the
arrival directions of the highest energy cosmic rays and the
positions of nearby objects is discussed elsewhere
(\cite{Hague:2009, Auger:Anisotropy}). Events above 55 EeV were
selected and 17 out of 44 correlate with the position of nearby
objects from the V\'eron-Cetty and V\'eron (VCV) catalog. The
cumulative binomial probability that an isotropic flux leads to 17
or more correlations is low, 0.006. Still, this correlation is
weaker than the one published in an earlier analysis
\cite{Auger:Science}.

\paragraph{Shower Depth of Maximum}
The analyses of the $X_{max}$ and rms($X_{max}$) measured values
are shown and discussed in \cite{AugerXmax:2010}. They support the
hypothesis that the transition from galactic to extragalactic
cosmic rays occurs in the ankle region. Moreover and assuming that
the hadronic interaction properties do not change much within the
observed energy range, both the observed $X_{max}$ and their
fluctuations independently are a signature of an increasing
average mass of the primary particles with energy up to 59 EeV,
which is of interest to analyze in contrast to the anisotropy
results (though in a higher energy region) since the latter tends
to imply a lighter composition, i.e. protons. Heavier primaries
will have a larger deflection in the galactic and intergalactic
magnetic fields, thus preventing the correlation with point
sources. Maybe the hadronic interactions change in this energy
range and as such further studies might cast light on them.

\paragraph{Photon Limits}

No high-energy photons were identified and the derived upper
limits are discussed in \cite{Auger:PhotonLimit}. The results
complement the previous constraints on top-down models from Auger
surface detector data. In future photon searches, the separation
power between photons and nuclear primaries can be enhanced by
adding information for the Auger Observatories upgrades (see
below).

\paragraph{Tau Neutrino Limit}
Data was analyzed to present an upper limit to the diffuse flux of
$\nu_{\tau}$ \cite{Auger:Neutrinos}, no neutrinos were detected
yet although the Auger Observatory has the best detection
sensitivity currently available around a few EeV, which is the
most relevant energy to explore the predicted fluxes of GZK
neutrinos. However in the worst case of systematic uncertainties,
the limit presented here is still higher by about one order of
magnitude than GZK neutrino predictions and as such more data
needs to be collected to make a final assessment.

\section{Upgrades of the Pierre Auger Observatory}
It is of uppermost relevance to study the 0.1 - 10
EeV energy range in order to cast light on the second knee and
ankle features. The two main requirements are good energy
resolution and primary type identification (statistical
discrimination over this energy range will suffice) since as
mentioned above the transition from galactic (heavier elements) to
extragalactic (lighter elements) sources is directly linked to
primary composition.

The two shower parameters relevant to composition are the
atmospheric depth at shower maximum, X$_{max}$, and the shower
muon content. Composition is very poorly understood in this energy
range where varieties of mixed compositions are reported ranging
from proton to iron dominated primaries (see \cite{ANC04} and
references within). Still, composition can only be assessed within
a given hadronic interaction model and therefore comparisons are
only to be performed under those premises. Moreover, they would
need a recalculation if a hadronic model is reformulated. Much
more robust results are attained from the variation rate of either
X$_{max}$ (called elongation rate) or muon content as a function
of energy \cite{ABU00} by which composition changes may be
assessed fairly independent of the assumed hadronic model.

The Auger Observatory is being upgraded to study the primary
particle type in the second knee - ankle region with both
fluorescence telescopes and muon counters (MC) giving the air
shower longitudinal profiles and muon contents, respectively. It
will therefore perform spectrum and composition measurements with
unprecedented accuracy.

Within the original Auger baseline design, the surface array is
fully efficient above $ \sim$ 3 EeV and in the hybrid mode this
range is extended down to $ \sim$ 1 EeV. There are three
enhancements to make the Observatory fully efficient down $\sim$
0.1 EeV: HEAT (\emph{High Elevation Auger Telescopes})
\cite{Klages:2007, Kleifges:2009}, AMIGA (\emph{Auger Muons and
Infill for the Ground Array}) \cite{Etchegoyen:2007, Platino:2009,
Buchholz:2009, Medina-Tanco:2009}, and AERA (\emph{Auger
Engineering Radio Array}) \cite{vdBerg:2009}. Moreover, these
enhancements focuss on composition analyses.

\paragraph{HEAT}

\begin{figure}[!ht]
  \centering
  \includegraphics[width=.40\textwidth]{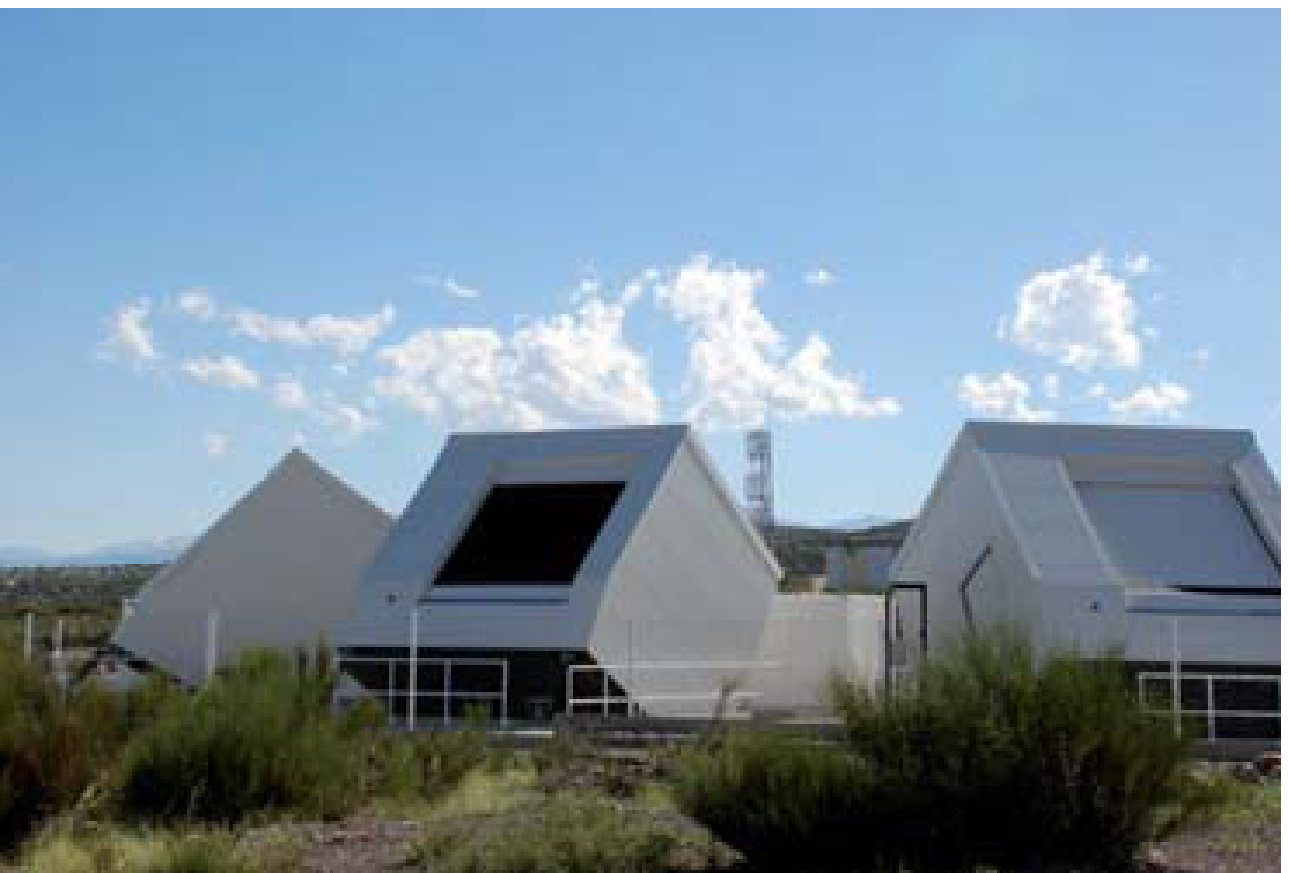}~\hfill
  \includegraphics[width=.35\textwidth]{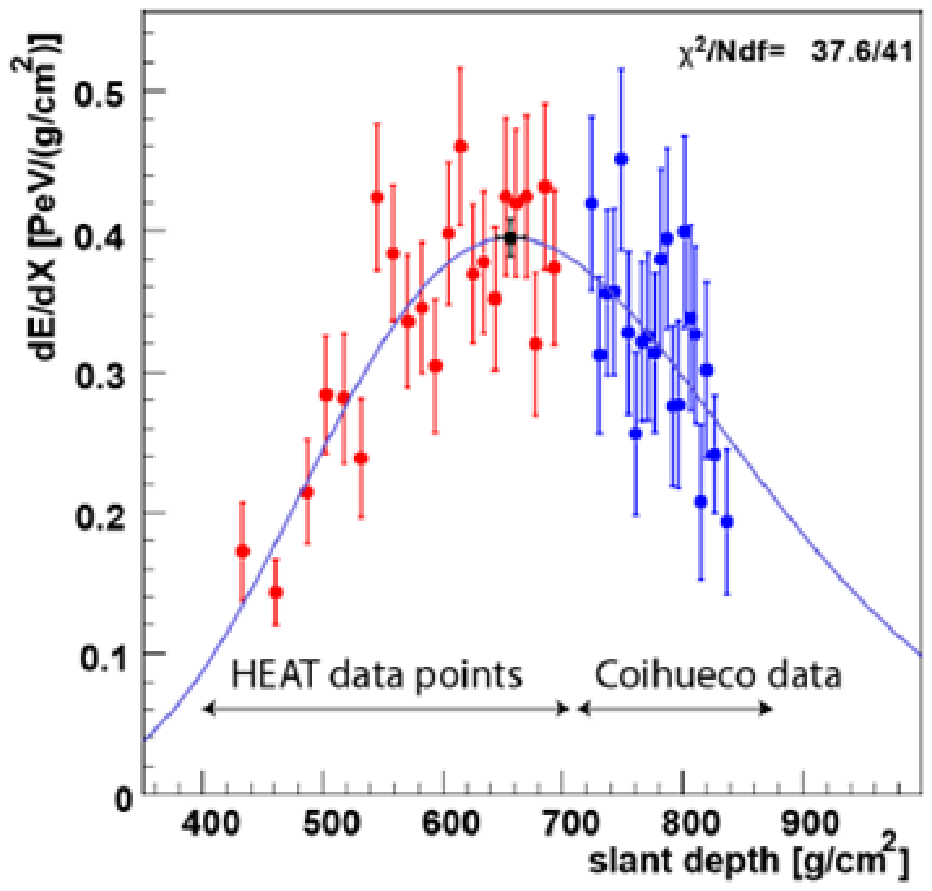}
  \caption{{\it (left)}  The three buildings of the HEAT telescopes at Cerro Coihueco;
  {\it (right)} longitudinal shower profile where both HEAT and Coihueco telescopes are
  needed in order to reconstruct the profile, event energy $(2.0 \pm 0.2) 10^{17}$ eV.}
  \label{fig:heat}\end{figure}

HEAT adds to the Observatory three new telescopes of similar
design to the previous ones in order to fully detect longitudinal
profiles of showers in the $1.5^{\circ }\times 58^{\circ }$
elevation range. The original telescopes cover the range of
$1.5^{\circ }\times 30^{\circ }$ while the HEAT telescopes
$30^{\circ }\times 58^{\circ }$. This is attained by tilting the
telescopes enclosures upwards by 29$^{\circ}$ (see Fig.
\ref{fig:heat}.left). This higher elevation FOV is needed for the
detection of lower-energy showers since they would develop earlier
in the atmosphere and they have to be detected at closer distances
because the fluorescence light produced is roughly proportional to
the primary energy. HEAT is already taking data and a
reconstructed longitudinal profile of a low energy event is
displayed in Fig. \ref{fig:heat}.right.

HEAT is optimized to record nearby showers in combination with the
existing telescopes at Coihueco as well as to take data with
AMIGA. The first measurements have showed that HEAT will improve
the energy threshold down to 0.1 EeV and that the operation
fulfills the design requirements.

\begin{figure}[!h]
\includegraphics[height=.2\textheight]{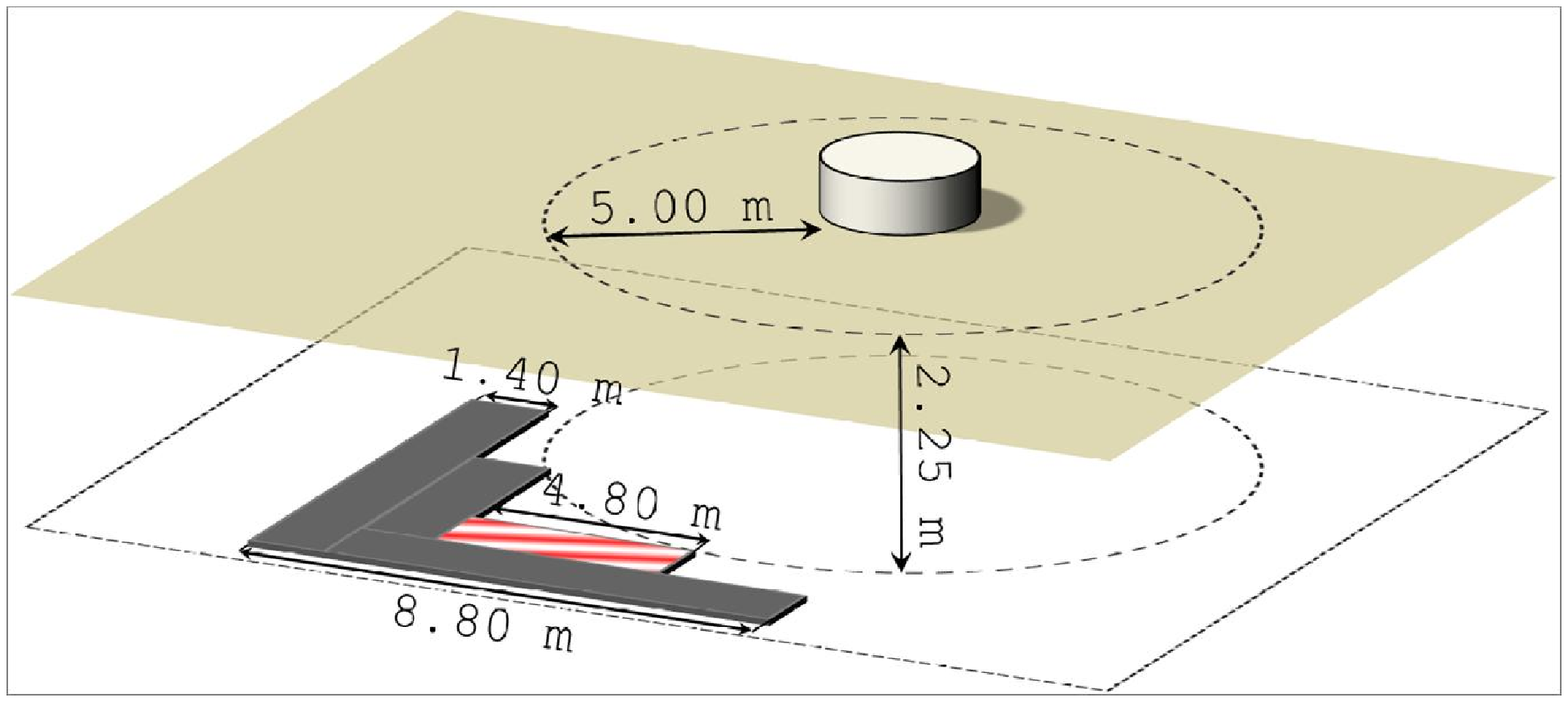}
\includegraphics [height=.2\textheight]{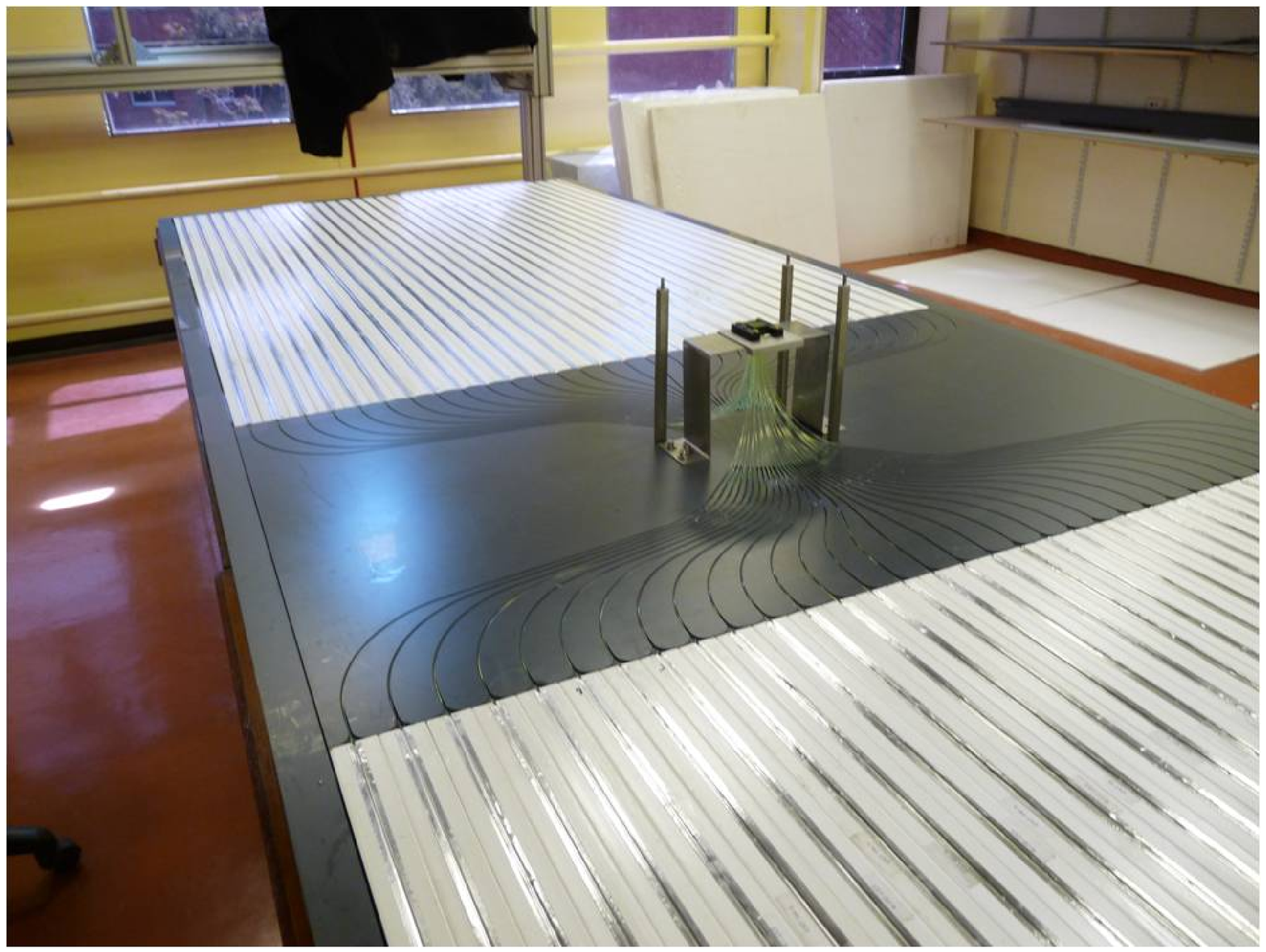}
\caption{ \emph{(left)} Layout of the SD-MC doublet, a 5 m$^{2}$
module depicted with red strips; \emph{(right)} first 5 m$^{2}$
prototype, scintillator strips layout and fiber routing onto the
optical connector. \label{Muon-counter}}
\end{figure}

\paragraph{AMIGA} AMIGA is being deployed over a small graded infilled
area of 23.5 km$^{2}$ (see Fig. \ref{Auger-map}.right) since the
cosmic ray flux rapidly increases as the energy threshold is
lowered. On the other hand, the detectors have to be deployed at
shorter distances among each other in a denser array since lower
energies imply smaller airshower footprints on the ground. A
graded infill of 433 and 750 m triangular grids was chosen in
order to optimize the detection over more than an order of
magnitude, from 3 EeV down to 0.1 EeV. Also, and since the two
main experimental requirements of the new detection system are
good energy resolution and primary type identification, AMIGA
consists of pairs of water Cherenkov surface detectors and muon
counters over viewed by FDs. It entails 85 of such pairs (Fig.
\ref{Muon-counter}.left).

Deployment begins with an engineering array called Unitary Cell, a
hexagon with 7 detector pairs, one in each hexagon vertex and one
in the center. These counters are composed of 4 modules each, 2
$\times$ 5 m$^{2}$ and 2 $\times$ 10 m$^{2}$ with 2 and 4 m long
strips, respectively. Each counter has an area of 30 m$^{2}$ and
it is made of 4 modules of 64 scintillator strips each, 32 on each
side of the PMT, (see Fig. \ref{Muon-counter}.right) with glued
optical fibers (doped with wave length shifters) in a groove
coated on top with a reflective foil. The strips are 1 cm thick
and 4.1 cm wide.

\begin{figure}[!ht]
  \centering
  \includegraphics[width=.45\textwidth]{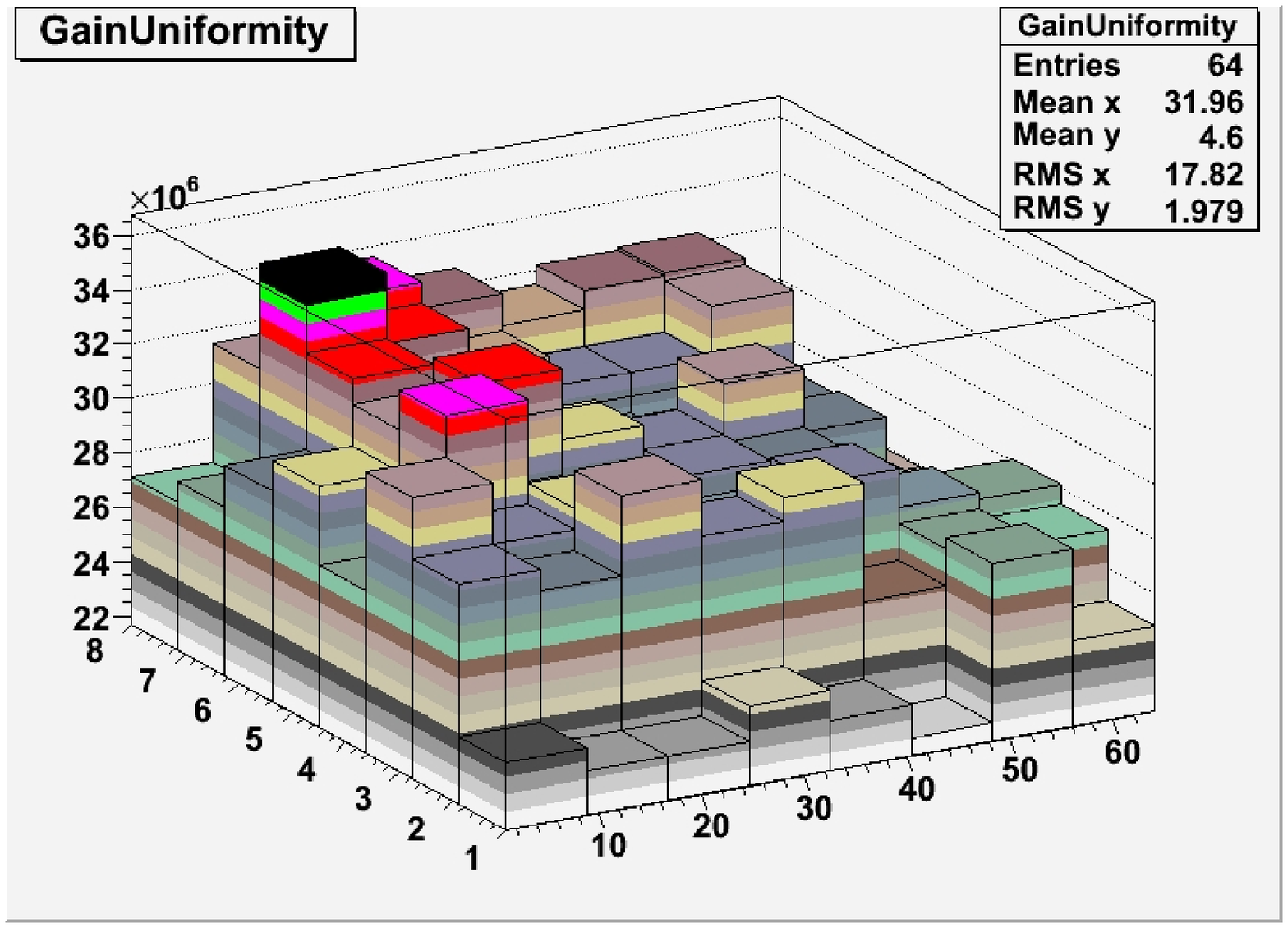}
  \includegraphics[width=.45\textwidth]{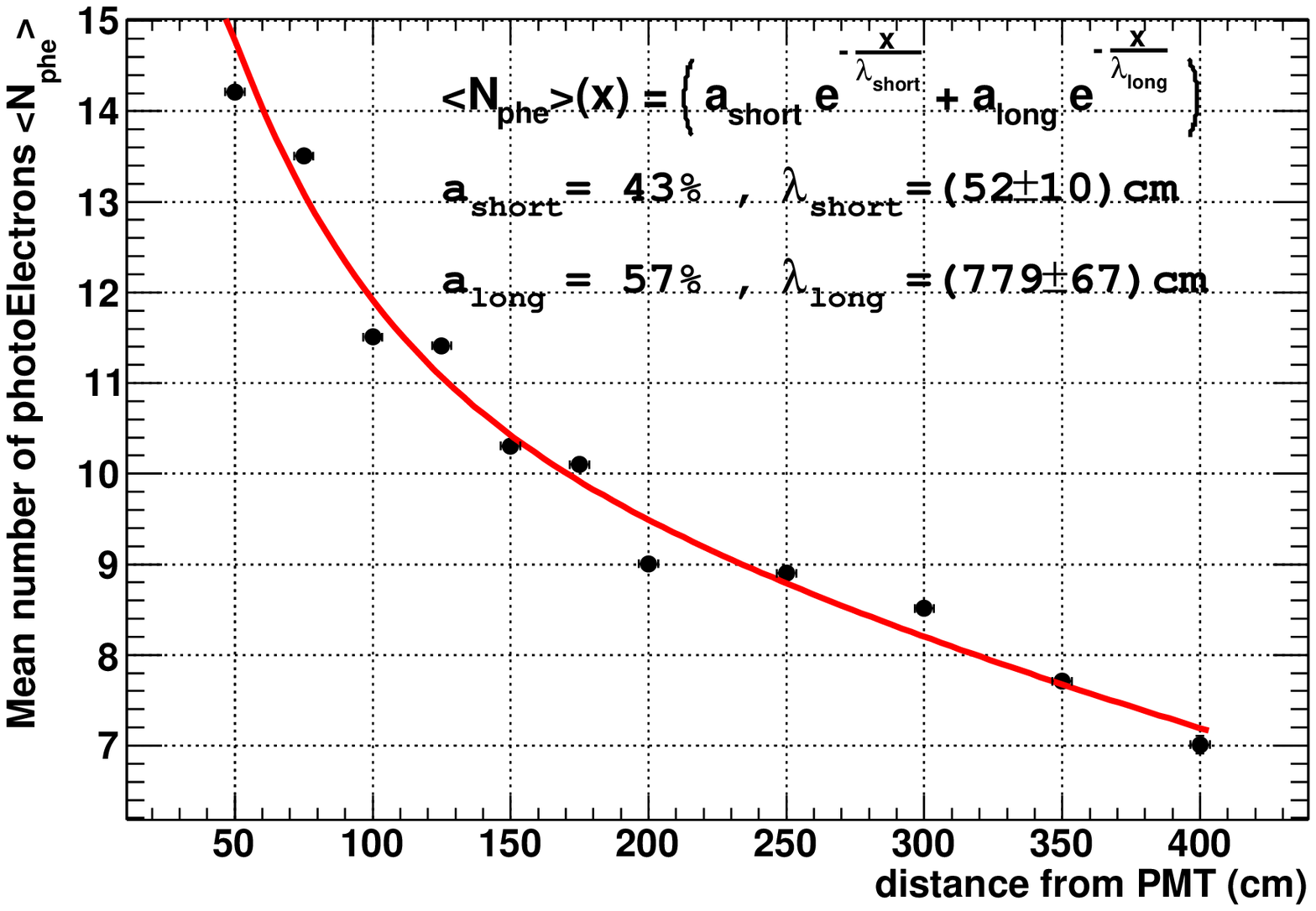}
  \caption{{\it (left)} gain uniformity of an 8 $\times$ 8 pixel Hamamatsu H8804MOD PMT;
  {\it (right)} light attenuation of a 1.2 mm Kuraray optical
fiber, light produced by background muons traversing a
scintillator strip.}
  \label{fig:fiberPMT}
\end{figure}

Each module has a 64 pixel high quantum efficiency Hamamatsu
H8804MOD PMT with a 2 mm $\times$ 2 mm pixel size. Since the PMT
is the counter sensitive element, its performance parameters need
to be known, assessed, and set (see e.g. Fig.
\ref{fig:fiberPMT}.left) in order to ensure its proper functioning
for data acquisition. Therefore, each AMIGA PMT will be analyzed
by a test facility designed and built for this purpose.

The front end bandwidth is of 180 MHz and the electronics sampling
is performed at 320 MSps (3.125 ns) with an external memory to
store up to 6 ms of data, equivalent to 1024 showers. The total
number of independent electronic channels per counter is 256. This
high segmentation requirement is an attempt to measure a single
muon per segment per unit time in order to avoid pile-up. A signal
is counted as a muon if it has two or more single photo electrons
(spe). In turn, an spe is registered when its amplitude is above
$\sim$ 30 \% of each pixel mean spe value. The counter so designed
is very robust and trustworthy because obtaining the muon number
from the integrated signal has significant disadvantages: i) the
number of spe per muon vary as much as a factor of $\sim$ 2 due to
fiber attenuation depending on whether the muon arrives at the
near or far end of the scintillator strip (see Fig.
\ref{fig:fiberPMT}.righ), ii) the light yield from the same fiber
type may vary a factor of $\sim$ 1.5 , and iii) changes in gain or
in spe numbers will impact on the muon counting via total charge
collected.

\begin{figure}[!ht]
  \centering
  \includegraphics[width=.45\textwidth]{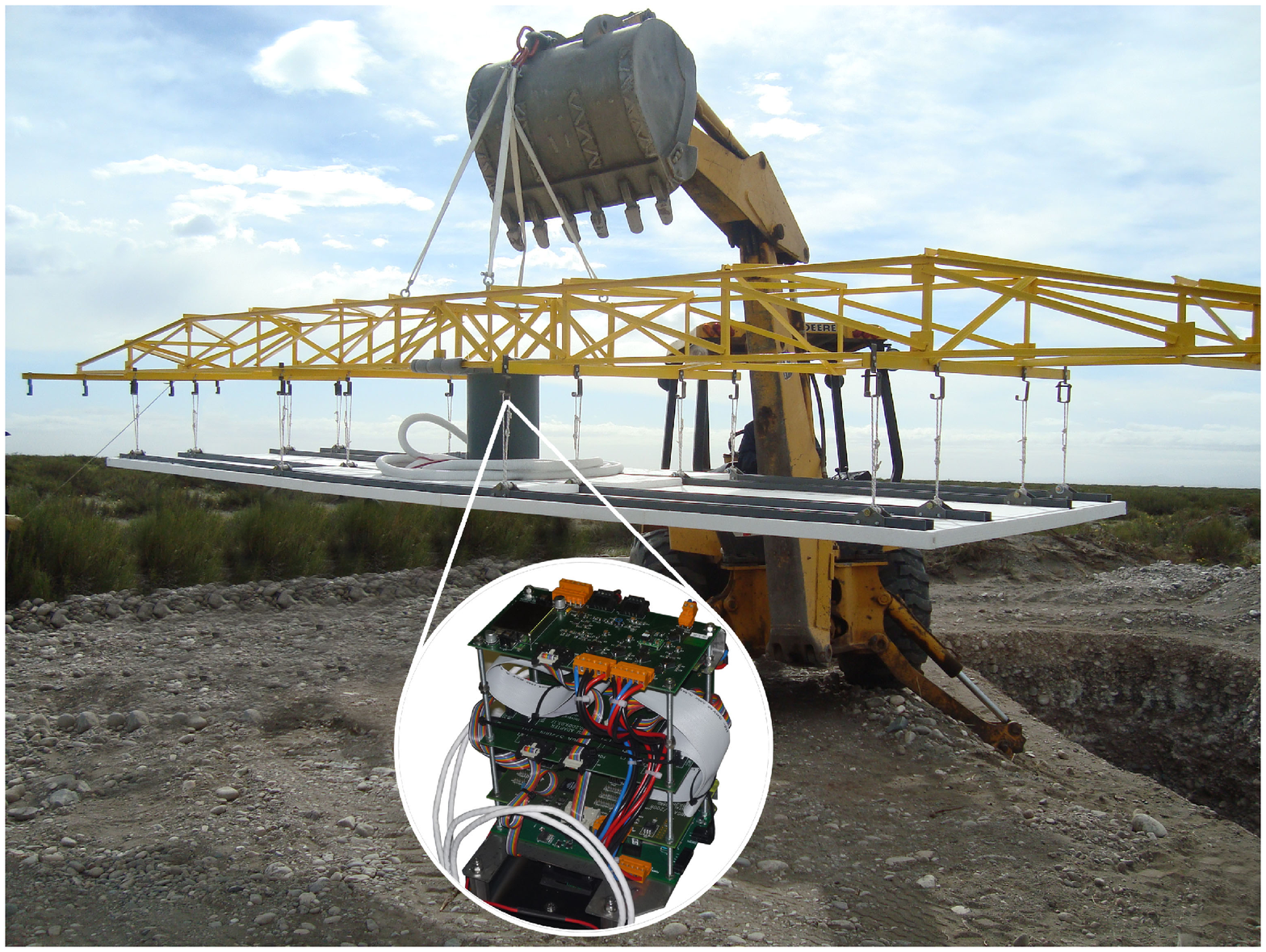}
  \includegraphics[width=.45\textwidth]{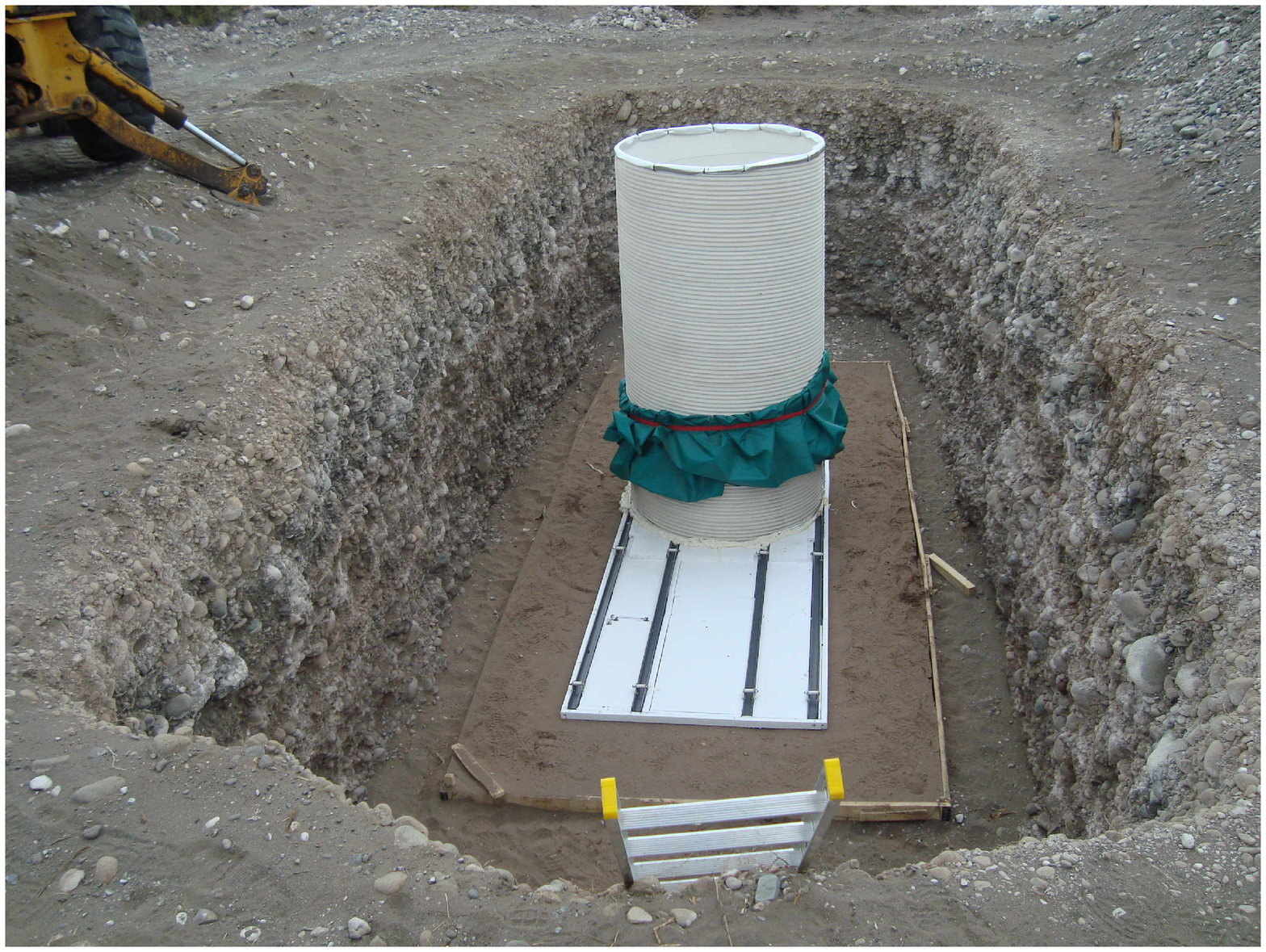}
  \caption{{\it (left)} First module buried at the Observatory site, the insert shows the electronics inside its
  enclosure; {\it (right)} module in the well placed over  a sand bed with service pipe
installed, about to be buried.}
  \label{fig:counter}
\end{figure}

The first fully equipped 5 m$^{2}$ prototype module has been
designed, built, tested, and buried at the Observatory site in
November 2009 (see Fig.\ref{fig:counter}), depicted in red strips
in Fig. \ref{Muon-counter}.left.

\begin{figure}[!ht]
  \centering
   \includegraphics[width=.45\textwidth]{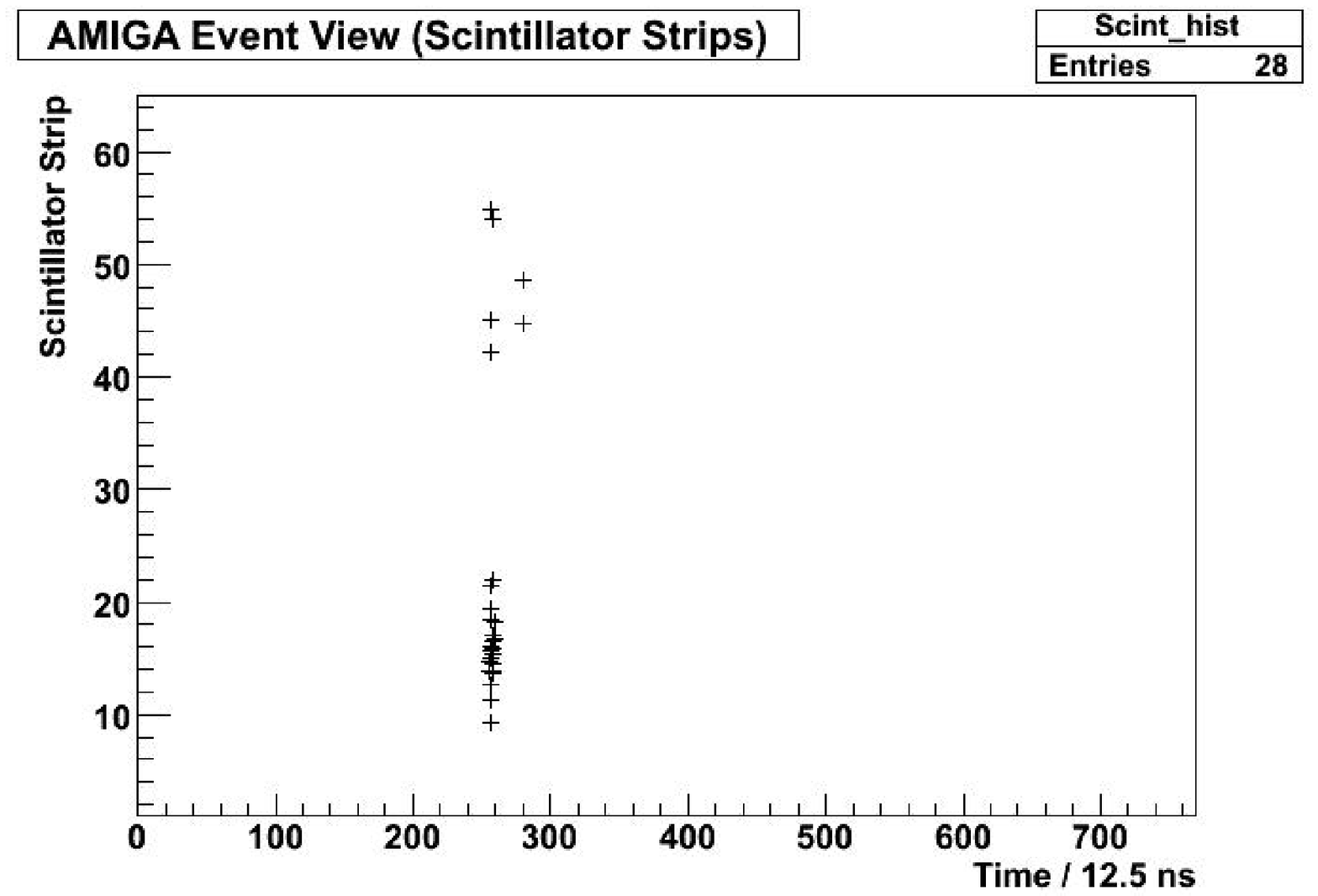}
   \includegraphics[width=.45\textwidth]{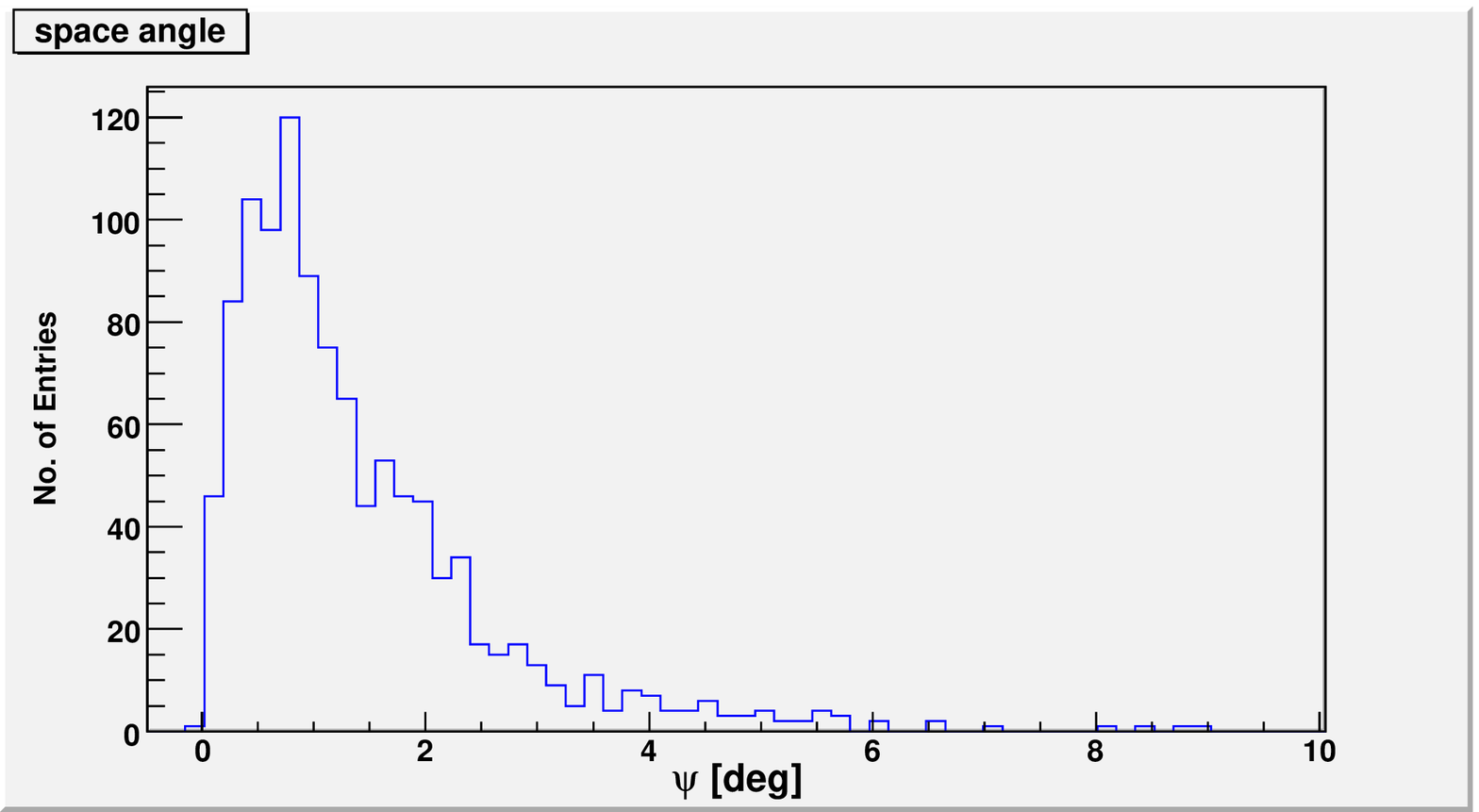}
  \caption{\it{(left)} A typical event recorded by the prototype module.
  On the horizontal axis is the time in units of 12.5 ns and
  on the vertical axis the scintillator strip number. Twelve strips triggered, all
  in the same time bin (\# 256); \it{(right)} histogram of the difference in the reconstruction of
arrival directions with
  and without the infill SDs.}
  \label{fig:space-angle}
\end{figure}

Data have been taken with this 5 m$^{2}$ prototype counter by
requiring at least 8 channels to simultaneously trigger in any
given time bin of 12.5 ns. A typical time structure of an event is
shown in Fig. \ref{fig:space-angle}.left.

Also 50/61 SDs of the 750 m infilled area are now operational and,
in particular, the 7 SDs of the unitary cell have been
instrumented with the new telecommunication system designed and
built for AMIGA (industrial grade radios XBee Pro working over IP
and using the IEEE 802.15.4 standard controlled with a local
TS7260 single board microcomputer). The graded infill was
envisaged to have both a saturated efficiency down to 0.1 EeV and
as a by-product to experimentally test the main array
reconstruction uncertainties. Preliminary tests have been
performed by the reconstruction of events with and without the
infill SDs with a data set with $\geq$ 7 and $\geq$ 3 or more SDs
for the infilled and main arrays, respectively \cite{Platino:2009,
Sidelnik:2009}. One of such a tests, a comparison of the arrival
directions, resulted in an upper limit 1.4$^{\circ}$ (see Fig.
\ref{fig:space-angle}.right). This shows the good performance of
the main array even with a reduced number of triggered stations
and below its 3 EeV lower energy threshold limit.

\paragraph{AERA}

AERA aims at the detection of cosmic-ray showers by measuring the
coherent radiation at radio frequencies emitted by secondary
particles, which are deflected by the geomagnetic field. The
radio-detection technique has been investigated already in the
1960's (see e.g. \cite{JEL65}) and results from more recent
experiments at energies beyond 10$^{17}$ eV (LOPES \cite{HAU09}
and CODALEMA \cite{LAU09}) show the great potential of this
technique. The main advantages of radio detection are a nearly
100\% duty cycle, a signal to noise ratio scaling with the square
of the cosmic-ray energy, and its high angular resolution and
sensitivity to the longitudinal air-shower evolution. These
features, combined with the capability of measuring the depth at
shower maximum (composition analysis) and the cost effectiveness
for instrumentation in large arrays, makes it an excellent
complement to the Auger SD array.

\begin{figure}[h!]
   \centering
   \resizebox{8cm}{!}{\includegraphics{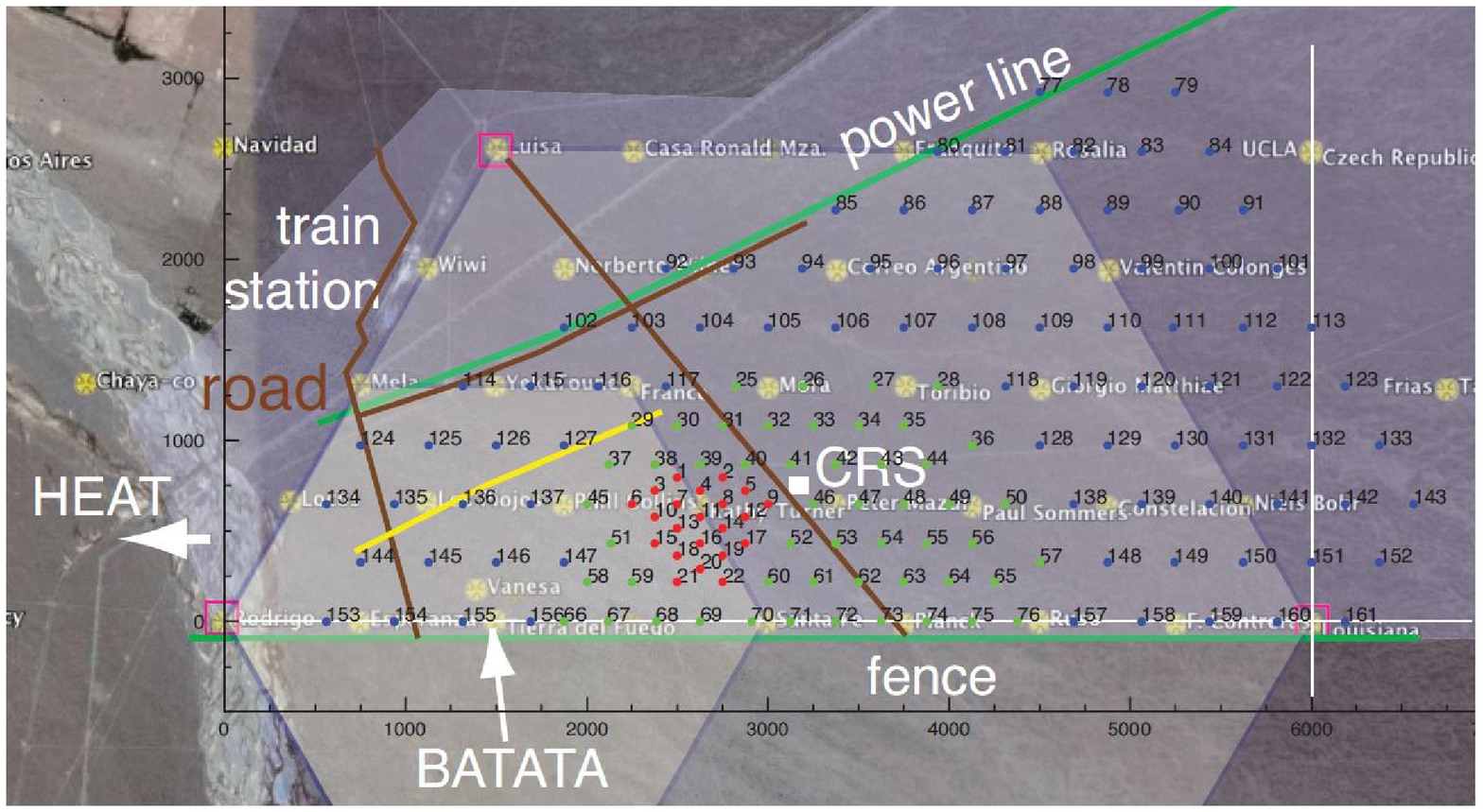}}
   \resizebox{6cm}{!}{\includegraphics{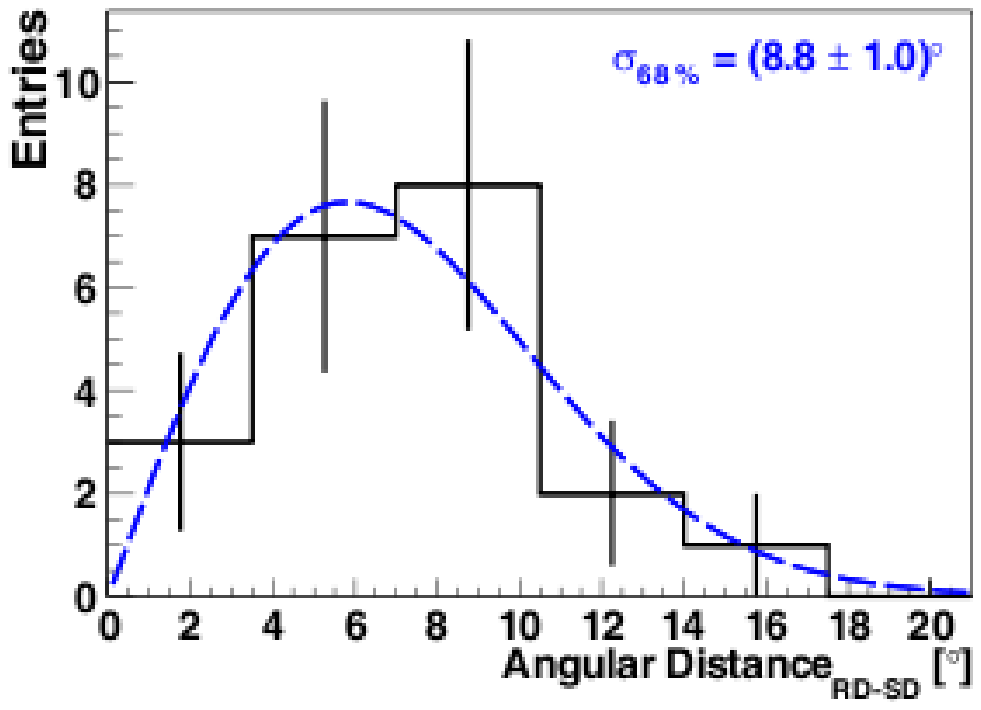}}
  \caption{\emph{(left)} Layout of the proposed AERA graded array overlaying the AMIGA
  hexagons in the background. The location of each antenna is marked as a red
  box;
 \emph{(right)} a preliminary comparison of reconstructed arrival directions with both SD and radio stations; the angular
resolution of these initial measurements was dominated by timing
resolution of the detector stations (see \cite{COP:09} for more
details).}
 \label{fig:aeramap}
\end{figure}

AERA is composed of 150 self-triggered radio-detection stations
over $\sim$ 20 km$^{2}$ \cite{vdBerg:2009} and it is now being
deployed in the AMIGA-HEAT region (see Fig.
\ref{fig:aeramap}.\emph{left}). It consists of a graded infilled
area with different station densities in order to cover a large
energy region above 0.1 EeV. Primary energies, types, and arrival
directions (see a preliminary result in Fig.
\ref{fig:aeramap}.\emph{right}) will be reconstructed by
super-hybrid measurements entailing four totally independent
measuring systems, SD, FD, MC, and radio-detection antennae.

In conclusion, the Auger Observatory has measured the high energy
cosmic ray spectrum and clearly identified the ankle and the
cutoff, it also found clear indications of anisotropy in the
arrival directions of cosmic rays with energies above 55 EeV.
Statistical composition analyses have been performed with depths
at shower maximum and their fluctuations which merit further
research particularly in hadron interactions at higher energies.
New upper limits have been found for photon and tau neutrino
fluxes. Enhancements are well under way in order to study the
transition region from galactic to extragalactic sources with
surface detectors, optical telescopes, muon counters and
radio-detection antennae with unitary efficiencies unbiased in
composition.

\bibliographystyle{aipproc}   

\end{document}